\documentclass[nofootinbib,showpacs,amsmath,amssymb]{revtex4}
\usepackage[latin1]{inputenc}
\usepackage{amsmath,amssymb}
\usepackage{latexsym}

\begin{document}
\title{Nonsmooth backgrounds in quantum field theory}
\author{M.~Bordag}\email{ Michael.Bordag@itp.uni-leopzig.de}
\author{D.~V.~Vassilevich}\email{Dmitri.Vassilevich@itp.uni-leipzig.de}
\thanks{
Also at V.A.Fock Institute of Physics, St.Petersburg University, Russia}
\affiliation{Institut f\"{u}r Theoretische Physik, Universit\"{a}t Leipzig, 
Augustusplatz 10, D-04109 Leipzig, Germany}

\preprint{LU-TP 2004/016}
\begin{abstract}
The one-loop renormalization in field theories can be formulated
in terms of the
heat kernel expansion. In this paper we calculate leading contributions
of discontinuities of background fields and their derivatives
to the heat kernel coefficients. These results are then used to
estimate contributions of the discontinuities to the Casimir
energy. Sign of such contribution is defined solely by the order
of discontinuous derivative.
We also discuss renormalization in the presence of singular
(delta-function) potentials. We show that an independent surface tension
counterterm is necessary. This observation seems to resolve some 
contradictions in previous calculations.
\end{abstract}
\pacs{11.10.-z, 02.40.-k}
\maketitle
\section{Introduction}
Recent years have seen much progress in understanding the Casimir
effect \cite{Milton01,Bordag:2001qi}. Although this effect is oftenly 
considered as a macroscopic manifestation of quantum fields, 
many ingredients of quantum field theory have not been yet 
reformulated to the Casimir framework. Renormalization is probably
the most important example of such an ingredient. 
Early calculations of the Casimir force between rigid bodies
were not affected by the renormalization problems since the divergent
part of the Casimir energy is local (does not depend on the distance)
and, therefore, does not contribute to the force. There are,
however, many quantities of interest (like the Casimir stress)
which do contain a divergent part and require some renormalization.

There exists a subtraction prescription \cite{Bordag:1998vs,Bordag:2001qi}
which allows to obtain 
well-defined results for the Casimir energy even if it is divergent.
This prescription consists in expanding the regularised Casimir energy
in log-power series for large mass $m$ of the fluctuating field and
subtracting all non-negative powers of $m$ and $\log m$ terms.
A similar prescription is being used in quantum field theory in
curved space \cite{Birrell:1982ix}. In two dimensional scalar theories
it is equivalent to the ``no tadpole'' condition \cite{Bordag:2002dg}.
Unfortunately, no such statement is known for other models or other
dimensions. Besides, this large mass prescription is not applicable
for massless fields.

In order to remove ambiguities in the Casimir energy, it was suggested
just a couple of years ago \cite{Graham:2002fi,Milton:2002vm} 
to perform the calculations in renormalizable theories. This direction
attracted a lot of interest, but the results of different groups are
still contradicting (cf.\ \cite{Jaffe:recent,Milton04} and references
therein).

The main difficulty in applying the renormalization procedure of quantum
field theory to the Casimir-type problems is that the latter one usually
assume singular backgrounds (which may be penetrable walls, or boundary
conditions, or just non-smooth potentials). This motivated us to study
quantum field theory (in the one-loop approximation) in the presence
of non-smooth background fields. Namely, we consider the case
when $p$th derivative of the background field jumps on a surface
$\Sigma$ of co-dimension one.

In this paper we use the zeta function regularization and the heat kernel
methods (both are sketched in sec.\ \ref{sreg}). In this regularization,
the one-loop divergences are expressed in terms of the heat kernel 
coefficients $a_k$. These coefficients are analysed in sec.\ \ref{shk}
for non-smooth potentials. In sec.\ \ref{shk1} we calculate the linear 
order in the potential and find how the localised heat kernel is
modified by the presence of singularities. Somewhat surprisingly,
the global heat kernel (which defines the counterterms) is not
sensitive to the singularities in this order. The quadratic order in
the potential is studied in sec.\ \ref{shk2}, where we calculate 
leading contributions to global heat kernel. There we prove a
``folk theorem'' that if standard ``smooth'' expression for $a_k$ is
divergent due to the singularity, a non-zero surface contribution
to $a_{k-1}$ should appear. These results allow us to calculate 
the leading contribution of the singularity to the Casimir energy
(sec.\ \ref{Casdens}). 
Then, in sec.\ \ref{sren} we discuss
renormalization of theories with delta-function singularities.
\section{The regularization}\label{sreg}
In this section we sketch basic technical results on the zeta-function
regularization and the heat kernel. For a more detailed introduction
one may consult the monographs \cite{mono} or the review
\cite{Vassilevich:2003xt}. The reader may also consult a recent paper by
Fulling \cite{Fulling:2003zx} which deals with applications of the
heat kernel technique to the Casimir energy calculations.

To simplify our discussion in this paper we consider scalar field
theories only. The classical action reads
\begin{equation}
\mathcal{L} = \int_M d^n x \left( \frac 12 (\nabla \varphi)^2 +
U(\varphi )\right) \,. \label{clac}
\end{equation}
We suppose that the manifold $M$ is flat, so that it is either torus
or $\mathbb{R}^n$. For $M=\mathbb{R}^n$ one has to assume some
fall-off conditions on the fields to obtain convergent integrals.
We assume also that the metric on $M$ has Euclidean signature.
We use the background field formalism, so that we split 
$\varphi =\Phi + \phi$ where $\Phi$ is a background field, and
$\phi$ is a quantum fluctuation.
Note, that the background field $\Phi$ may also include 
a part which describes ``external conditions'' (like, e.g., boundary
conditions or domain walls or membranes). To calculate the one-loop
effective action one should expand the action (\ref{clac}) about
the background field $\Phi$ keeping the quadratic order terms 
in fluctuations only.
\begin{equation}
\mathcal{L}_2= \frac 12 \int_M d^n x\, \phi D[\Phi] \phi \,,
\label{L2}
\end{equation}
where $D[\Phi]$ is an operator of Laplace type,
\begin{equation}
D=-(\nabla^2 + E),\qquad E=-U''(\Phi )\,. \label{oper}
\end{equation}
For a multicomponent $\varphi$ the potential $E$ is matrix-valued.

Formal path integration over $\phi$ leads to the following result
for the one-loop effective action
\begin{equation}
W=\frac 12 \ln \det (D) \,.\label{Wdet}
\end{equation}
Right hand side of (\ref{Wdet}) is divergent and has to be regularised.
To this end we use an integral representation for the determinant
\begin{equation}
W_s = -\frac 12 \mu^{2s} \int_0^\infty \frac {dt}{t^{1-s}} K(t,D)\,,
\label{Ws}
\end{equation}
with a regularization parameter $s$ (the regularization is removed
in the limit $s\to 0$). The heat kernel $K(t,D)$ is defined as
a functional trace,
\begin{equation}
K(t,D)=\mathrm{Tr}_{L^2} (e^{-tD})\,. \label{KtD}
\end{equation}
Let us define the zeta function of $D$ by the equation:
\begin{equation}
\zeta (s,D)= \mathrm{Tr}_{L^2} (D^{-s}) \,.\label{defze}
\end{equation}
Now we can rewrite $W_s$ in terms of the zeta function
\begin{equation}
W_s = -\frac 12 \mu^{2s} \Gamma (s) \zeta (s,D) \,.\label{Wsz}
\end{equation}
This procedure is called the zeta-function regularization
\cite{zetafreg}.

The regularised effective action
(\ref{Wsz}) has a pole at $s=0$:
\begin{equation}
W_s=-\frac 12 \left( \frac 1s -\gamma_E +\ln \mu^2 \right)
\zeta (0,D) - \frac 12 \zeta'(0,D) \,,\label{Wzpole}
\end{equation}
where $\gamma_E$ is the Euler constant.

There is an asymptotic series as $t\to +0$
\begin{equation}
K(f,t,D)=
\mathrm{Tr}_{L^2} (f\exp (-tD)) 
\cong\sum_{k\ge0}t^{(k-n)/2}a_k(f,D)\,.
\label{asymptotex}
\end{equation}
(here we have introduced a heat kernel smeared with a smooth function
$f$. It is related to the unsmeared kernel (\ref{KtD}) by
$K(t,D)=K(1,t,D)$.) If $E$ is smooth and $M$ has no boundary,
all odd numbered coefficients in (\ref{asymptotex}) vanish,
$a_{2k+1}=0$. 

There is an important relation
\begin{equation}
\zeta (0,D) = a_n (1,D) \label{zan}
\end{equation}
which tells that the one-loop divergences are defined by the heat
kernel asymptotics. We remind that $n$ is dimension of the underlying
manifold $M$.
Besides that, the heat kernel expansion defines
short-distance behaviour of the propagator and the large mass expansion
of the effective action.

We shall use also a bi-local kernel $K(x,y;t)$ of $e^{-tD}$. It is related
to the smeared heat kernel by means of the equation
\begin{equation}
K(f,D,t)=\int_M dx\, \mathrm{tr} f(x) K(x,x;t) \,.\label{bilocla}
\end{equation}
Here $\mathrm{tr}$ denotes a matrix trace over all discrete indices
if they are in the model. It should be distinguished from $\mathrm{Tr}_{L^2}$
which is a functional trace in the space of square integrable functions.

Let us consider a surface $\Sigma$ of co-dimension one in $M$.
We take a background field $\Phi$ which
is smooth everywhere except
for $\Sigma$ where its' $p$th normal derivative has a discontinuity.
The potential $E$ shall then possess the same property. At least
locally, $\Sigma$ divides $M$ into two parts $M^+$ and $M^-$.
Let $\nu^+$ and $\nu^-$ be unit normal vectors to $\Sigma$
pointing inside $M^+$ and $M^-$ respectively. Since $M$ is smooth,
$\nu^-=-\nu^+$, but this may be not true for more general geometries.
Let $E^{(k)+}$ (respectively $E^{(k)-}$) be a limit of $k$th 
derivative w.r.t. $\nu^+$ (resp. $\nu^-$) of $E(x)$ as $x\to \Sigma$
from the $M^+$ (resp. $M^-$) side. If the $k$th derivative is continuous,
$E^{(k)+}=(-1)^kE^{(k)-}$. For example, if $M=\mathbb{R}^1$ and 
$\Sigma=\{x=0\}$, then $M^\pm=\mathbb{R}_\pm$, $E^{(1)\pm}=\pm \partial_x E$.
\section{Heat kernel for non-smooth potentials}\label{shk}
In this section we restrict ourselves to the case of flat $\Sigma$.
For a non-flat $\Sigma$ the heat kernel expansion should contain
additional terms with extrinsic curvature of $\Sigma$. Such terms
have larger canonical (mass) dimensions than the terms considered
below and shall, therefore, contribute to higher heat kernel coefficients.
\subsection{Local heat kernel in the linear order}\label{shk1}
To analyse the heat kernel asymptotics we use
the perturbative expansion  \cite{Barvinsky:uw}
(see also \cite{Vassilevich:2003xt}
for a short overview).  The exponent
$\exp (-tD)=\exp (t(\Delta + E))$ with $\Delta =\partial_\mu^2$
can be expanded in a power series
in $E$:
\begin{equation}
e^{-tD}=e^{t\Delta} + \int\limits_0^t ds e^{(t-s)\Delta} E e^{s\Delta}
+\int\limits_0^t ds_2 \int\limits_0^{s_2} ds_1
e^{(t-s_2)\Delta}E e^{(s_2-s_1)\Delta} E e^{s_1\Delta} +\dots
\label{pertex}
\end{equation} 
This expansion is purely algebraic. 
Each order of $E$ in (\ref{pertex}) is given by a convergent
integral if $E$ is smooth or has a singularity located on
a surface of co-dimension one (this can be a discontinuity
of a derivative or even a delta-function singularity).
In the case of $\delta$-singularities on a submanifold
of co-dimension two or higher, there might be problems with the
convergence. More careful estimates can be found in Ref.\ 
\cite{Bordag:1999ed}.

Here we analyse the heat kernel expansion to the linear order in
$E$. Let $f$ be a smooth function on $M$. Let $K_0(x,y,t)$ be the
heat kernel for $E=0$. Then to this order
\begin{equation}
K(f,E,t)=K(f,0,t) - \int_0^t d\tau \int_M dx \int_M dy f(x)
K_0(x,y,\tau -t) E(y) K_0(y,x,\tau )+\dots \label{linord}
\end{equation}
Obviously, the second term in (\ref{linord}) is symmetric w.r.t.
exchanging the role of $E$ and $f$. Therefore, it can be
interpreted as a linear order term of the heat kernel with the
potential $f$ and the smearing function $E$. This term then reads:
\begin{equation}
\int_M dx \tilde K_1(x,x,t) E(x)\label{2ndterm}
\end{equation}
where $\tilde K_1$ is the heat kernel with the potential $f(x)$ at
the linear order in $f$. The crucial point is that $f$ is smooth.
If we neglect all curvatures, then local heat kernel coefficients
corresponding to $\tilde K$ are:
\begin{equation}
\tilde a_{2k}(x) =(4\pi )^{-n/2} \alpha_2(k) \Delta^{k-1} f(x)
\label{a2k}
\end{equation}
where \cite{Gilkey:1979}:
\begin{equation}
\alpha_2(k)=\frac {2k!}{(2k)!} \label{alpha2}
\end{equation}
The heat kernel coefficients we are interested in read:
\begin{equation}
a_{2k}=  (4\pi )^{-n/2} \alpha_2(k) \int_M dx E(x)\Delta^{k-1}
f(x) \label{hk2k}
\end{equation}
Odd-numbered coefficients are all zero. Formula (\ref{hk2k})
can be checked for $k=1,2$ \cite{Gilkey:2001mj}.

An interesting observation regarding (\ref{hk2k}) is that all
these coefficients vanish for $f=1$, i.e. one loop divergences
and the large mass expansion
in the effective action are not affected by non-smoothness
of the potential (to the linear order studied in this section).

\subsection{Global heat kernel to quadratic order}\label{shk2}
In this section we study the heat trace asymptotics in the order
$E^2$ and prove for this case a statement which existed
in the folklore for many years: If due to a singularity of the
background the volume contribution to $a_{2k}$ diverges,
the coefficient $a_{2k-1}$ should have a non-zero surface
contribution. This statement is based on particular case calculations 
for scalar backgrounds \cite{Bordag:fv,Bordag:1998vs} and for
singular magnetic fields \cite{Drozdov:xw}, and on analytic
calculations of \cite{Bordag:1999ed,Gilkey:2001mj}. A similar 
conclusion for dielectric problems follows from the analysis
of \cite{Bordag:1998vs}. In the present paper we give explicit
expressions for the leading contribution to the heat kernel
expansion from discontinuities of derivatives of the potential
of arbitrary order.

We start with analysing relevant volume and surface invariants.
For a smooth potential $E$ there is an invariant $(\nabla^{p+1} E)^2$
(which means $(\nabla^\mu \Delta^{p/2} E)^2$ for $p$ even and
$(\Delta^{(p+1)/2} E)^2$ for $p$ odd).
It has dimension $(2p+6)$ and can contribute
to the coefficient $a_{2p+6}$:
\begin{equation}
a_{2p+6} \simeq (4\pi)^{-n/2} \beta (p) {\rm tr} \int_M dx
(\nabla^{p+1} E)^2 \,.\label{a2p6}
\end{equation}
If $p$th derivative of $E$ is discontinuous the expression (\ref{a2p6})
contains a delta-function squared and is, therefore, meaningless.
Then we expect that the following heat kernel coefficient appears
\begin{equation}
a_{2p+5} \simeq (4\pi)^{-(n-1)/2} \gamma (p) {\rm tr} \int_\Sigma dx 
(E^{(p)+}-(-1)^pE^{(p)-})^2\,. \label{a2p5}
\end{equation}
No other invariant of the same dimension can appear. For example,
$(E^{(p)+}-(-1)^pE^{(p)-})(E^{(p)+}+(-1)^pE^{(p)-})$ changes sign
if one exchanges the role $M^-$ and $M^-$, 
$(E^{(p)+}+(-1)^pE^{(p)-})^2$ would give a non-zero $a_{2p+5}$ 
even for smooth potentials.
According to the general theory both $\beta (p)$ and $\gamma (p)$
do not depend on $n$. We have to make sure that they are non-zero.

We use again the perturbative expansion (\ref{pertex}). 
The heat trace can be also expanded,
\begin{equation}
K(t,D)={\rm Tr}\left( e^{-tD}\right)=
\sum_{j=0}^\infty K_j(t),\label{pexhtr}
\end{equation}
where $K_j$ contains the $j$th power of $E$. Localised version of
$K_1$ has been studied above. In the next order we have:
\begin{eqnarray}
K_2(t)=&&{\rm Tr} \left( \int\limits_0^t ds_2 \int\limits_0^{s_2} ds_1
e^{(t-s_2)\Delta}E e^{(s_2-s_1)\Delta} E e^{s_1\Delta} \right) \nonumber \\
=&&{\rm tr} \int_M dy \int_M dz 
\int\limits_0^t ds_2 \int\limits_0^{s_2} ds_1 
K_0 (z,y;t-s_2+s_1)E(y)\nonumber\\
&&\qquad\qquad\qquad\qquad\qquad \times K_0(y,z;s_2-s_1)E(z) \,.\label{K2a}
\end{eqnarray}
To derive (\ref{K2a}) we used cyclic property of the functional
trace in order to combine $e^{(t-s_2)\Delta}$ with $e^{s_1\Delta}$.

Since the coefficients $\beta (p)$ and $\gamma (p)$ are universal constants
we can use some particularly simple model to calculate their values.
We take $M=\mathbb{R}$, $E(x<0)=0$, $E(x)$ smooth for $x>0$ and decreases
exponentially fast as $x\to +\infty$. We also suppose that first $p-1$
derivatives are continuous at $x=0$. Then 
\begin{equation}
K_0(x,y;t)=(4\pi t)^{-1/2} \exp \left( -\frac{(x-y)^2}{4t} \right)
\end{equation}
We adopt the strategy of \cite{Marachevsky:2003zb}. After removing a
redundant integration (\ref{K2a}) takes the form:
\begin{equation}
K_2(t)=\frac t{8\pi} \int_0^\infty dx \int_0^\infty dy \int_0^t d\tau
\frac{\exp \left[ -(x-y)^2 \left( \frac 1{4\tau} + \frac 1{4(t-\tau)} 
\right) \right]}{\sqrt{\tau (t-\tau)}} E(x) E(y) \,.\label{K2b}
\end{equation}
Next we integrate over $\tau$ to obtain
\begin{equation}
K_2(x,y;t)= \frac t8 \int_0^\infty dx \int_0^\infty dy \,
\mathrm{erfc} \left[ \frac{|x-y|}{\sqrt{t}} \right]
E(x) E(y) \,.\label{K2c}
\end{equation}
Let us change the variables
\begin{eqnarray}
&&x=z+r\sqrt{t} ,\qquad y=z \quad {\mbox{for}}\ x>y \nonumber\\
&&y=z+r\sqrt{t} ,\qquad x=z \quad {\mbox{for}}\  x<y .
\end{eqnarray}
In both cases $k, r \in [0, +\infty[$. Then
\begin{equation}
K_2(x,y;t)= \frac {t^{3/2}}4 \int_0^\infty dz \int_0^\infty dr \,
\mathrm{erfc} (r) E(z) E(z+r\sqrt{t}) \label{K2d}
\end{equation}
The $t\to 0$ asymptotic expansion is now performed by using the 
following formula:
\begin{equation}
\int_0^\infty f(r\sqrt t)  \mathrm{erfc} (r) \simeq \sum_{n=0}^\infty
t^{n/2}
\frac{\Gamma \left( 1+\frac n2 \right) }{(n+1)! \sqrt{\pi}} f^{(n)}(0)
\label{erfas}
\end{equation}
Now we only have to pick up relevant terms in the expansion. The
term with $n=2(p+1)$ reads:
\begin{equation}
\frac{t^{p+3}}{(4\pi t)^{1/2}} \frac 12 \frac{2k!}{(2k)!}
\int_0^\infty dz E(z) \partial_z^{2(p+1)} E(z) \,,\label{eventerm}
\end{equation}
which is consistent with (\ref{a2p6}) for smooth potentials if
\begin{equation}
\beta (p) =(-1)^{p+1} \frac 12 \alpha_2 (p+2)\,.\label{gb1}
\end{equation}
$\alpha_2$ is given  by (\ref{alpha2}) above. Equation (\ref{gb1})
follows easily from \cite{Gilkey:1979} and can be used as
a consistency check.

The term with $n=2p+1$ has the form:
\begin{equation}
t^{p+2} 4^{-(p+2)} \int_0^\infty dz E(z) 
\partial_z^{2p+1} E(z) \,.\label{oddterm}
\end{equation}
If first $p-1$ derivatives of $E$ vanish at $z=0$, this result is consistent
with (\ref{a2p5}) and gives
\begin{equation}
\gamma_p = (-1)^{p+1} 2^{-2p-5} \,.\label{gb}
\end{equation}
We see, that both constants $\gamma$ and $\beta$ are non-zero.

Our calculation also confirms that as long as general
``smooth'' formulae for the heat kernel coefficients give convergent
integrals no modfications appear due to the singularities.
In the present context this means that old non-modified formulae
are valid for $a_k$ with $k<2p+5$.
\subsection{Application: Casimir energy density at the singularity} 
\label{Casdens}
Let us calculate leading contribution of non-smoothness of the
potential to the vacuum energy. In the zeta function
regularization the ground state energy is defined as
\begin{equation}
\mathcal{E} (s) =\frac 12 \mu^{2s}\sum_k \varepsilon_k^{1-2s}
\,,\label{E0s}
\end{equation}
where the regularization parameter $s$ should be taken zero after
the calculations, $\epsilon_k$ are eigenfrequencies of elementary
excitations defined as square root of eigenvalues of the
Hamiltonian:
\begin{equation}
H=-\partial_i^2 + V(x) + m^2=D + m^2 \,.\label{Ham}
\end{equation}

We shall work in a $3+1$ dimensional theory. Therefore, the
operator $D$ is three dimensional. $V$ is a static potential which
which is supposed to be non-negative. We can rewrite (\ref{E0s})
though the zeta function of $H$,
\begin{equation}
\mathcal{E} (s) =\frac 12 \mu^{2s} \textrm{Tr} \left(H^{\frac 12
-s} \right)=\frac 12 \mu^{2s} \zeta \left({s-\frac 12}, H \right)
\label{Eze}
\end{equation}
The zeta function, in turn, can be expressed through the heat
kernel:
\begin{equation}
\zeta \left(s-\frac 12, H \right) = \frac 1{\Gamma \left( s
-\frac 12 \right)} \int _0^\infty dt \, t^{s- 3/2} K(t,H)
\label{zHK}
\end{equation}
and
\begin{equation}
K(t,H)=K(t,D) e^{-tm^2} \,.\label{KHKD}
\end{equation}

It is easy to see that the large mass expansion of the vacuum
energy is generated by the small $t$ asymptotics of $K(t,D)$.
Indeed, by substituting the heat kernel expansion in (\ref{Eze})
with (\ref{zHK}) and integrating over $t$ we obtain:
\begin{equation}
\mathcal{E} (s) =\frac 12 \mu^{2s}\frac 1{\Gamma \left( s -\frac
12 \right)}  \sum_{k} a_k(D) m^{-2s -k +4} \Gamma \left( s+\frac
k2 -2 \right)\,. \label{Eexp}
\end{equation}
As expected, the terms with $k=0,\dots ,4$ are divergent in the limit
$s\to 0$.

We would like to separate contributions from non-smooth parts 
of the potential. If $p$th derivative jumps,
the leading contribution comes from the coefficient $a_{2p+5}$ in the
heat kernel expansion. Note, that in four dimensions all contributions
of these type are not divergent, so that we can put $s=0$ already in
(\ref{Eexp}). Then the term we are interested in reads
\begin{equation}
\mathcal{E}^{[p]}= -\frac 1{4\sqrt{\pi}} a_{2p+5} m^{-2p-1} \Gamma \left(
p+\frac 12 \right) \,.\label{Ep1}
\end{equation}
Next we use (\ref{a2p5}) and (\ref{gb}) to obtain
\begin{equation}
\mathcal{E}^{[p]}= (-1)^{p} \frac {(2p-1)!!}{4\pi} 2^{-3p-7}
m^{-2p-1} \int_\Sigma dx\ \left( \delta V^{(p)} \right)^2 \,,
\label{Ep2}
\end{equation}
where $\delta V^{(p)}$ is discontinuity of of $p$th normal derivative of
$V$ on $\Sigma$, according to our conventions $(-1)!!:=1$.

We see, that discontinuities in the potential itself and
in its' even order derivatives tend to increase the vacuum energy,
while discontinuities in odd order derivatives tend to decrease the vacuum
energy. Of course, practically it may be not easy to separate
contributions from continuous and discontinuous parts. 

\section{Renormalization with singularities}\label{sren}
As we have already seen in the presence of singularities the heat
kernel expansion is modified. This means that new counterterms may appear
in quantum field theory at one loop. Non-smoothness of the potential
modifies global heat kernel coefficients starting with $a_5$ (discontinuous
potentials) or even higher (discontinuous derivatives). Therefore, they
have no effect on counterterms in four dimensions. In order to be closer
to physical applications we consider a stronger singularity 
(delta-potentials) in four dimensions. This example is of particular
interest because of extensive discussion in the literature (cf.
recent works \cite{Jaffe:recent,Milton04} and references therein). 

We start with some technical information. Consider an operator
which has a singular part concentrated on the surface $\Sigma$:
\begin{equation}
D=-(\partial^2 + E(x) + v(x) \delta_\Sigma (x))\,. \label{Dwithdel}
\end{equation}
For simplicity, we do not consider here any gauge fields and suppose
that $E(x)$ is smooth. A mathematically correct formulation of
spectral problem for operators with delta-like singularities 
yields a spectral problem for the regular part of $D$ outside
of $\Sigma$ supplemented by matching conditions on $\Sigma$.
\begin{eqnarray}
&&\phi^+\vert_\Sigma = \phi^{-}\Sigma \,,\nonumber\\
&&\left[\phi^{(1)+} + \phi^{(1)-} - v\phi \right]_\Sigma =0 \,.
\label{mcond}
\end{eqnarray}
We remind, that according to our notations $\phi^{(k)\pm}$ is
a $k$th normal derivative of $\phi$ calculated on $M^+$ or
$M^-$ side of $\Sigma$. For a smooth $\phi$: 
$\phi^{(1)+}=-\phi^{(1)-}$ (cf. sec.\ \ref{sreg}).

The heat kernel coefficient $a_4$ which is responsible for one-loop
divergences in four dimensions reads \cite{Bordag:1999ed}:
\begin{equation}
a_4(D) = (4\pi)^{-n/2} \mathrm{tr} \left[ \int_M dx \frac 12 E^2
+\int_\Sigma dx \left( \frac  16 v^3 + E v \right) \right]
\label{a4delta}
\end{equation}
Note, that no additional terms appear in (\ref{a4delta}) even if the
surface $\Sigma$ is curved. This implies that one needs the same 
counterterms on spherical surfaces as on flat ones.

Consider now an action containing a surface interaction
term:
\begin{equation}
\mathcal{L}=\int_M d^4x \left( \frac 12 (\nabla\varphi)^2 +
\frac{m^2}2 \varphi^2 +\frac \lambda{12} \varphi^4 +\Lambda \right)
+\int_\Sigma d^3x \left( \sigma + \frac{\bar\lambda}2 \varphi^2 \right)\,.
\label{1stact}
\end{equation}
Here $m$, $\lambda$, $\Lambda$, $\sigma$, and $\bar\lambda$ are some
coupling constants.
Next we introduce a continuous\footnote{A stronger requirement
that $\Phi$ is smooth would exclude classical solution of this model.}
background field $\Phi$, $\varphi=\Phi+\phi$ and expand
the action (\ref{1stact}) up to quadratic order in quantum fluctuations
$\phi$. As a result, we obtain (\ref{L2}) where $D$ is given by
(\ref{Dwithdel}) and
\begin{equation}
E=-m^2 -\lambda \Phi^2,\qquad v=-\lambda\,.\label{Ev}
\end{equation}

With the help of (\ref{Wzpole}), (\ref{zan}), (\ref{a4delta}) and
(\ref{Ev}) we immediately obtain that the one loop divergences can be
cancelled by the following counterterms:
\begin{eqnarray}
&&\delta\Lambda = \frac 1s \frac{m^4}{32\pi^2} + \mathcal{O}(s^0),
\qquad
\delta m^2 = \frac 1s \frac{\lambda m^2}{16\pi^2} + \mathcal{O}(s^0),
\qquad
\delta\lambda = \frac 1s \frac{3\lambda^2}{16\pi^2} + \mathcal{O}(s^0),
\nonumber \\
&&\delta \sigma = \frac 1s \frac{1}{32\pi^2} \left( -\frac 16\bar\lambda^3
+\bar\lambda m^2 \right)+ \mathcal{O}(s^0),\qquad
\delta\bar\lambda = \frac 1s \frac{\bar\lambda \lambda}{32\pi^2} 
+ \mathcal{O}(s^0),
\label{ct}
\end{eqnarray}
We see that the model (\ref{1stact}) is renormalizable at least at one
loop.

It is important that both constants $\Lambda$ (the cosmological constant)
and $\sigma$ (the surface tension) must be present in the action to achieve 
renormalizability. One can consistently put $\Lambda =0$ to all 
orders of perturbation theory since its' \textit{observed} value
is negligible. The surface tension $\sigma$ cannot be excluded on the
same grounds. It has to be considered as an (experimental) input, as
other coupling constants. To remove all ambiguities in the model 
one needs five independent
normalisation conditions (including the ``trivial''
one $\Lambda =0$) which should fix the $\mathcal{O}(s^0)$ terms
in (\ref{ct}). To fix $\delta m^2$ and $\delta\lambda$ one can proceed as
in non-singular theories. Namely, one can consider the effective
action for a slowly varying background field $\Phi$
located far away from $\Sigma$. Then $\delta m^2$ and $\lambda$ can be
determined by prescribing certain values to the second and fourth
derivatives of the effective action w.r.t. $\Phi$ at $\Phi=0$.
Physically this is equivalent to fixing position of the pole in
the propagator of bosons and to fixing the value of the four-boson
vertex for zero external momenta (both processes have to be considered
very far from $\Sigma$). In principle, finite renormalization
of $\bar\lambda$ can be fixed by relating the renormalized
value of $\bar\lambda$ to some amplitudes of scattering of $\varphi$
on $\Sigma$. It is not clear however which condition is the most
convenient one. 
Unless all these conditions have been formulated one
cannot give a scheme-independent meaning to the surface tension.
This provides an alternative explanation to the contradictions
reported in \cite{Milton04}. We should also note, that if one is
interested only in the one-loop Casimir energy and stress 
on the background with $\Phi=0$ (as in \cite{Milton04}), 
ambiguities in finite renormalization
of $m^2$, $\lambda$ and $\bar\lambda$ are not important since they
are of the order $\hbar$ and do not enter the classical energy
for $\Phi=0$.

It is instructive to compare (\ref{1stact}) to other models which
appeared in the literature. The model considered by Milton \cite{Milton04}
did not contain the self-interaction $\lambda$ (which is not important)
and the surface tension $\sigma$ (which is very important
for discussing the counterterms). The model of the MIT group 
\cite{Jaffe:recent} uses coupling to an external source. Although, the
counterterm action suggested in \cite{Jaffe:recent} indeed allows to
remove all divergences (at least as long as the source is smooth), the
model shares a common difficulty of all models with non-dynamical
external fields: the latter either have to be fixed artificially or 
require infinite number of ``experiments'' to be properly determined
(if one ``experiment'' allows to determine one real number).
Indeed, each coefficient in the Taylor expansion of the external
source (or each form-factor) plays the role of an independent coupling.

There is a somewhat exotic example of a model with a surface interaction,
\begin{equation}
\widetilde{\mathcal{L}}=\int_M d^4x \left( \frac 12 (\nabla\varphi)^2 
+\frac \lambda{12} \varphi^4 \right)
+\int_\Sigma d^3x\, \frac{g}2 \varphi^3 \,,
\label{2ndact}
\end{equation}
which does not require any surface tension counterterms in the 
zeta-regularization at one loop. In other schemes (momentum cut-off, e.g.)
such counterterms may appear also for (\ref{2ndact}).

\section{Conclusions}\label{conlcu}
In this paper we studied quantum field theory on non-smooth backgrounds.
We have analysed leading contributions of the singularities to
the heat kernel coefficients.
We found that influence of the non-smoothness is rather mild:
no new counterterms appear in dimension up to 4. We were also able to prove
an important statement on the behaviour of the heat kernel expansion
for singular backgrounds: ``smooth'' formulae for the heat kernel
coefficients are valid as long as they are convergent. If the ``smooth''
expression for
$a_{2k}$ diverges for some $k$, a non-zero surface term in $a_{2k-1}$
inevitably appears.   
We demonstrated that discontinuities of
even order derivatives contribute a positive amount to the Casimir
energy, while contributions from discontinuities of odd order
derivatives are negative.

We have also analysed renormalization of theories with stronger
(delta-function) singularities. Our main message is that an independent
counterterm to the surface tension is needed in this case. Therefore,
the one loop surface tension crucially depends on the normalisation
condition. If no such condition is specified, one cannot give a
scheme-independent meaning to the surface tension.

One may extend our results by using the
heat kernel expansion with non-smooth gauge fields \cite{Moss:2000gv},
non-smooth geometries \cite{Gilkey:2001mj}, other surface singularities
\cite{Gilkey:2002nv}, and, perhaps, even singularities of conical type
\cite{Bordag:1996fw}.
\begin{acknowledgments}
This work was supported by the DFG project BO 1112/12-1.
\end{acknowledgments}

\end{document}